\documentclass[useAMS,usenatbib, twocolumn ,times, english]{mn2e}
\usepackage{times}
\usepackage[T1]{fontenc}
\usepackage[latin1]{inputenc}
\usepackage{verbatim}
\usepackage{color}
\usepackage{graphicx}
\usepackage{amssymb}
\providecommand{\boldsymbol}[1]{\mbox{\boldmath $#1$}}

\definecolor{Blue}{rgb}{0,0.08,0.65}
\definecolor{Red}{rgb}{0.65,0.08,0.05}
\definecolor{Green}{rgb}{0.15,0.45,0.25}

\def\mathfrak#1{{\mathrm{#1}}}

       \def\be{{\bf e}}

\def\be{\begin{equation}}
\def\ee{\end{equation}}
\def\ba{\begin{eqnarray}}
\def\ea{\end{eqnarray}}

\usepackage{babel}
\makeatother

\begin{document}

\title[On the Onset of Stochasticity in $\Lambda$CDM  Cosmological Simulations]{On the Onset of Stochasticity in $\Lambda$CDM  Cosmological Simulations}
\author[J. Thiebaut, C. Pichon , T. Sousbie, S. Prunet \& D. Pogosyan]{J. Thiebaut$^{1}$, C. Pichon$^{1,2}$ , T. Sousbie$^{1}$,  S. Prunet$^{1}$ \&  D. Pogosyan$^{3}$ \\
 $^{1}$Institut d'astrophysique
de Paris \& UPMC (UMR 7095), 98, bis boulevard Arago , 75 014, Paris,
France.\\
$^{2}$ Service d'Astrophysique, IRFU,(UMR)  CEA-CNRS, L'orme des meurisiers, 91
470, Gif sur Yvette,
France.\\
$^{3}$  Department of Physics, University of Alberta, 11322-89 Avenue, Edmonton, Alberta, T6G 2G7, Canada  
}
\maketitle
\begin{abstract}
The onset of stochasticity is measured in $\Lambda$CDM cosmological
simulations using a set of classical observables. It is quantified as
the local derivative of the logarithm of the dispersion of a given
observable (within a set of different simulations differing weakly
through their initial realization), with respect to the cosmic growth
factor. In an Eulerian framework, it is shown here that chaos appears
at small scales, where dynamic is non-linear, while it vanishes at
larger scales, allowing the computation of a critical transition scale
corresponding to $\sim 3.5 $Mpc$/h$.\\ This picture is confirmed by
Lagrangian measurements which show that the distribution of
substructures within clusters is partially sensitive to initial
conditions, with a critical mass upper bound scaling roughly like the
perturbation's amplitude to the power $0.15$.  The corresponding
characteristic mass, $M_{\rm crit}=2 \, 10^{13 } M_{\odot}$, is
roughly of the order of the critical mass of non linearities at $z=1$
and accounts for the decoupling induced by the dark energy triggered
acceleration.

The sensitivity to detailed initial conditions spills to some of the
overall physical properties of the host halo (spin and velocity
dispersion tensor orientation) while other ``global'' properties are
quite robust and show no chaos (mass, spin parameter, connexity and
center of mass position).  This apparent discrepancy may reflect the
fact that quantities which are integrals over particles rapidly
average out details of difference in orbits, while the other
observables are more sensitive to the detailed environment of forming
halos and reflect the non-linear scale coupling characterizing the
environments of halos.
\end{abstract}
\section{Introduction}\label{sec:Introduction}
%

Concerns regarding the predictability of cosmological measurements in
simulations have been with us for some time. In the neighbouring field
of secular galactic evolution, it has been known for a while
(\cite{sellwood}) that the significant undersampling of resonances
could mislead the dynamical evolution of N-body systems when the
evolution time becomes large compared to the local dynamical
time. Over the course of the last decade, various {}``universal''
relationships (\cite{nfw}, \cite{fctmasse}, \cite{fctmasse2}) have
been extracted numerically from cosmological N-body simulations. In
this context, significant efforts (\cite{power}) have been invested in
comparing different numerical schemes and codes, but, with the
development of very high resolution {}``zoom'' re-simulations
( \cite{eke},  \cite{Diemand}, \cite{moore} \cite{steinmetz}, \cite{vialactea}) one question
remains: how sensitive is a given run with respect to its initial
conditions?  In particular, what set of observables is likely to be
robust with respect to a specific choice in the ``phases'' of the draw
(the whitened initial realization)?  In the context of cosmology, the
general assumption has been that, even though the detailed orbits of
dark matter particles are likely to be poorly resolved by the numerical
schemes implemented, the properties of structures would nevertheless
be well represented \textsl{statistically}.  {In other words, so long
as the simulated region was large enough to represent a fair sample of
dynamically independent regions, the stochastic exponential departure
from the unperturbed trajectories was expected to average out when
considering such a statistical sample.}  The question remains for
features specific to a given realization, such as the relative
position of objects.

The sensitivity of the gravitational N-body problem to small changes
in initial conditions has been investigated in details by Kandrup and
collaborators in a series of papers
(\cite{Kandrup91,Kandrup92a,Kandrup92b,Kandrup94}) in the context of a
Newtonian (time-independent) Hamiltonian. They have shown in
particular that the growth of small perturbations in initial
conditions is exponential, with a mean e-folding time that is
asymptotically independent of the number of particles at large N, and
a distribution of e-folding times that is reproducible from simulation
to simulation for sufficiently large N. In the cosmological context,
the N-body description is an approximation of the collisionless
Boltzmann equation for the evolution of the dark matter, so that
another related question is in which sense a limit to the continuum
can be established as the number of particles increases.
Indeed, it has been argued in the literature
(\cite{Kandrup90,Goodman93}) that the discretization of smooth, and
possibly integrable potentials invariably leads to strongly chaotic
orbits in the N-body framework, independently of the number of
particles; this has been confirmed numerically
(\cite{Kandrup01,Sideris02}), both for integrable and non-integrable
underlying distributions by evolving orbits in {}``frozen-N body''
samplings of the smooth mass distributions.  However, Kandrup and
Sideris also showed that when one follows the deviation of orbits
evolved in the frozen-N and smooth potentials with identical initial
conditions, and when the deviation amplitude is allowed to reach large
fractions of the system size (macroscopic view), a continuum limit can
be well defined in the sense that these macroscopic departures from
the orbits of the smooth potentials (which can be themselves either
regular or chaotic) grow as a power law with time, and that the
characteristic divergence time is growing with the number of
particles.  These results, together with the claim of universality
(halo profiles (\cite{nfw}), their shape (\cite{moore2}), the mass
functions (\cite{fctmasse}, \cite{fctmasse2})) that is widely used in
the cosmology community, lead us to revisit the problem of the
sensitivity of N-body simulations to slight changes in the initial
conditions at fixed power spectrum in the cosmological context, and to
numerically investigate the presence (or absence) of ``chaotic''
behaviour of different statistical quantities derived from N body
simulations.  Our focus will be on the transition between large scale
linear dynamics and small scale stochastic properties (\cite{valuri}).
A possible concern in this context is the development of stochasticity
induced by the ill-conditioning/non-linearity of the estimator of the
chosen set of observables. Another concern lies in the specificities
of the numerical code used. Finally, numerical noise induced by
round-off errors should also be kept at bay, since they by themselves
will lead to some level of stochasticity. Since the topic of this
paper is not optimal estimation, no attempt will be made to argue that
the set of estimators used here is superior or offers a better
trade-off in bias versus variance. Similarly, a standard integrator
(\cite{gadget}) is used to carry the simulations with a set of
conservative parameters. Round-off errors are assumed to be
irrelevant.  {Specifically, this paper will investigate what scale and
mass is expected to play a role and will identify which quantities are
found to be robust with respect to such exponential divergence; it
will also find out if stochasticity breaks in as soon as non
linearities occur or if it is possible to identify two distinct time
scales in the dynamics of large scale structures.

This paper is organized as follows: in Section \ref{sec:Method} the method
to characterize the statistical onset of stochasticity in numerical N-body
simulations is presented. In Section \ref{sec:euler} the corresponding
Lyapunov exponents are computed for Eulerian quantities and (in Section \ref{sec:lagr})
for Lagrangian ones. Section \ref{sec:Conclusion} discusses other issues and wraps up.
\section{Method \& Settings}\label{sec:Method}
In this paper, we address the problem of the sensitivity of N-body
simulations to initial conditions. To do so, we choose to study how
slight changes in these initial conditions affect the evolution of the
dispersion of a number of statistical  quantities with time. This is
achieved by generating several realizations of identical simulations
where a small amount of random noise has been added to the initial realisation. 
The generic procedure goes as follows:
\begin{enumerate}
\item Generate a cosmological simulation using $grafic$
  (\cite{grafic}) and gadget (\cite{gadget}). 
\item Start with the same noise file (i.e. the ``phases''), but add a Gaussian
white noise with RMS 1/30$^{th}$ of the previous white noise (except
in Sections \ref{sub:trans} and \ref{sub:Spin-Orientation}, where
this amplitude is varied). This only affects the relative
positions of clumps, not their spectral distribution (the
expectation of the power spectrum remains unchanged).
\item Rerun the simulation with the new white noise;
\item Iterate the above two steps \textasciitilde{}50 times;
\item Compute a set of observables in each simulation;
\item Compute the RMS (or the relative RMS) of the distribution of observables
for various expansion factors.
\item Fit the corresponding evolution of the $\log$ RMS vs 
the expansion
factor.
\item Possibly find the scaling of its corresponding Lyapunov exponent
(see below), with the smoothing scale associated with the observable
(see Sec.\ref{sub:trans}), or the corresponding mass (see
Sec.\ref{sub:Spin-Orientation}).
\end{enumerate}
Let us define the ``Lyapunov exponent'', $\lambda_{X}$, as the rate of
change of the logarithm of the fluctuation of the relevant quantity,
$X$, as a function of the scale factor,
$a$:
\begin{equation}
\lambda_{X} \equiv \frac{{\rm d}\ln{\sigma_{X}}}{{\rm d }a}\,.
\end{equation}
 This stochasticity
parameter is not strictly speaking a Lyapunov exponent since it
corresponds neither to an asymptotic limit at large time, nor to an
asymptotic limit at small fluctuation.  It is closer in spirit to the
short time Lyapunov exponent defined by \cite{short}.

In practice two distinct sets of simulations are considered in this
paper, one composed of 65 realisations of $128^{3}$ particles each
($S_{1}$ hereafter) and the other of 27 realisations with $256^{3}$
particles each ($S_{2}$ hereafter). The box size is $100h^{-1}$Mpc,
the cosmology a standard $\Lambda$CDM model
($\Omega_{\rm m}=0.3,\,\Omega_{\Lambda}=0.7,\, H_{0}=70$), the softening
parameter is $39.5h^{-1}$kpc  and the expansion
factor ranges from 0.05 up to 1 for $S_{1}$ and from 0.05 up to 0.4
for $S_{2}$ .  These two sets allow us to check the robustness of our
finding with respect to resolution.  Lyapunov exponents will also be
expressed as characteristic timescales, $\tau$, using the relationship
between time and expansion factor in a CDM model (or equivalently in a
$\Lambda$CDM model below $a\leq0.5$), $a\propto\tau^{2/3}$.  Note that
the resolution in mass of FOF halos containing more than 100 particles
corresponds here to $4 \,10^{12} \, M_{\odot}$ for the set $S_{1}$ and
$5 \, 10^{11}\, M_{\odot}$ for the set $S_{2}$ .
\section{Eulerian exponents }\label{sec:euler}
%
In this Section, we investigate the ``global'' chaos in the evolution
of the Eulerian properties of the {\it density} field with respect to
the expansion factor $a$, as opposed to chaos in the Lagrangian
properties of {\it objects} which are specific to the matter
distribution in the universe (such as halos and filaments). This will
be addressed in Section \ref{sec:lagr}.

\subsection{Chaos in density fluctuations}\label{sub:density}
In order to study the density fluctuations, the density fields of
$S_{1}$ and $S_{2}$ are sampled on a $64^{3}$ grid using a simple NGP
(nearest grid point) method allowing the computation of statistical
quantities on the resulting grid, such as the average density or the
density fluctuations.\footnote{we also considered  a $128^{3}$ grid and found no difference in the measured exponents. }  Within each set, for every pixel we compute the
PDF of the realisations of the density values at that pixel and then
average the individual pixel PDFs over all the pixels that have mean
density value across realisations above the given threshold. The
evolution of the width of this pixel-averaged PDF is then computed as
a measure of the chaotic divergence amongst realisations with slightly
different initial conditions.
\begin{figure}
\includegraphics[width=0.8\columnwidth]{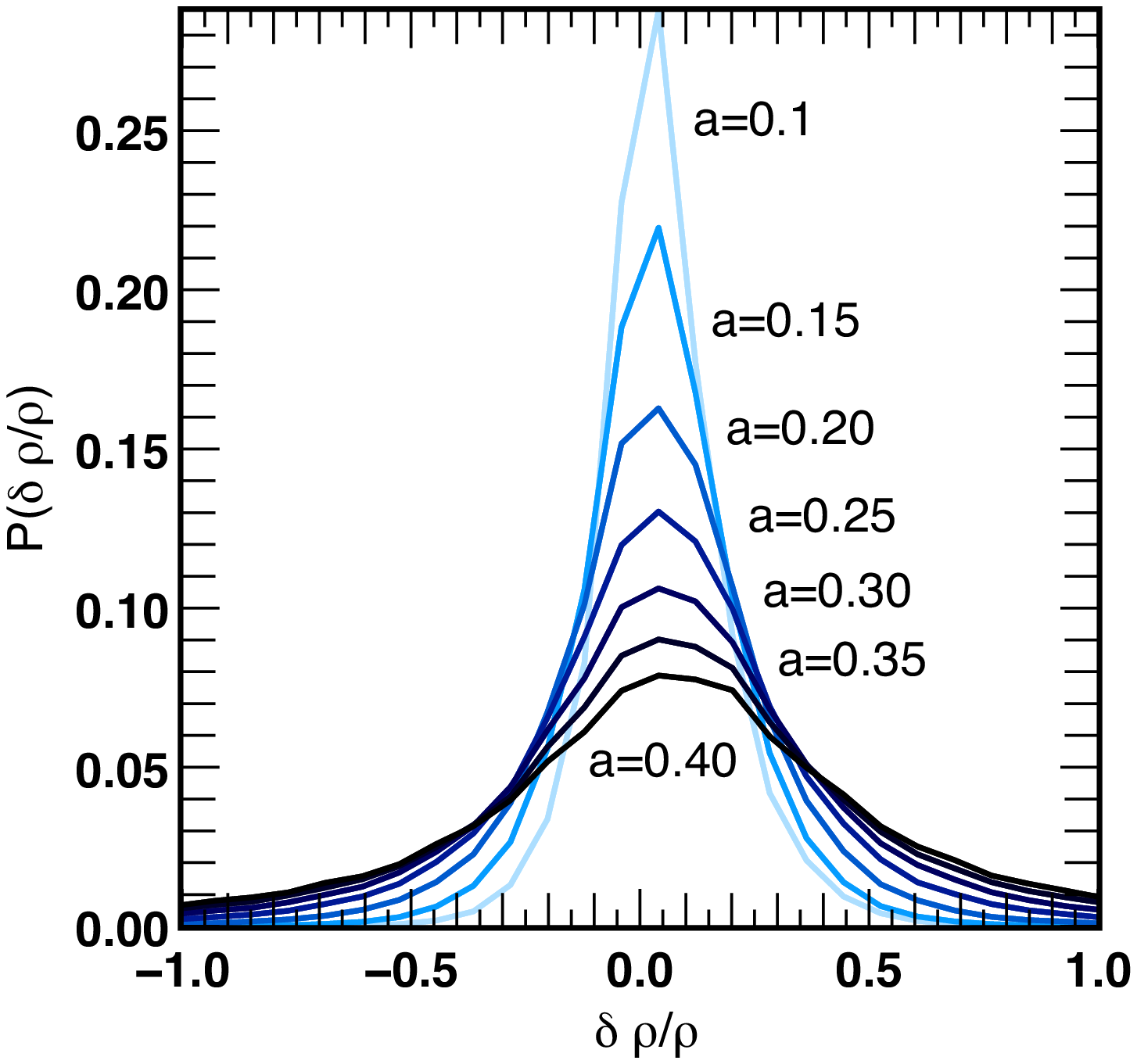}
\includegraphics[width=0.8\columnwidth]{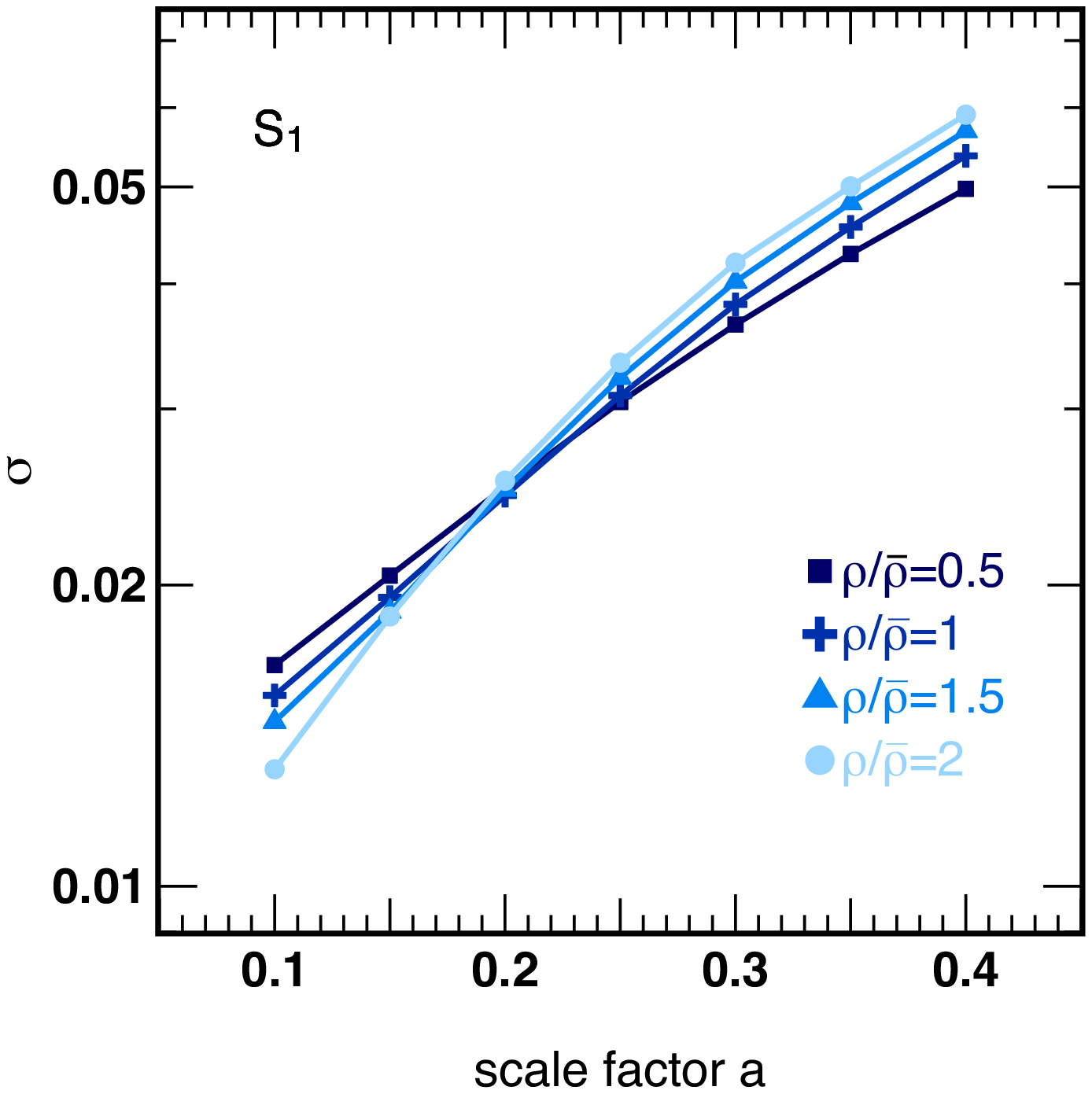}
\includegraphics[width=0.8\columnwidth]{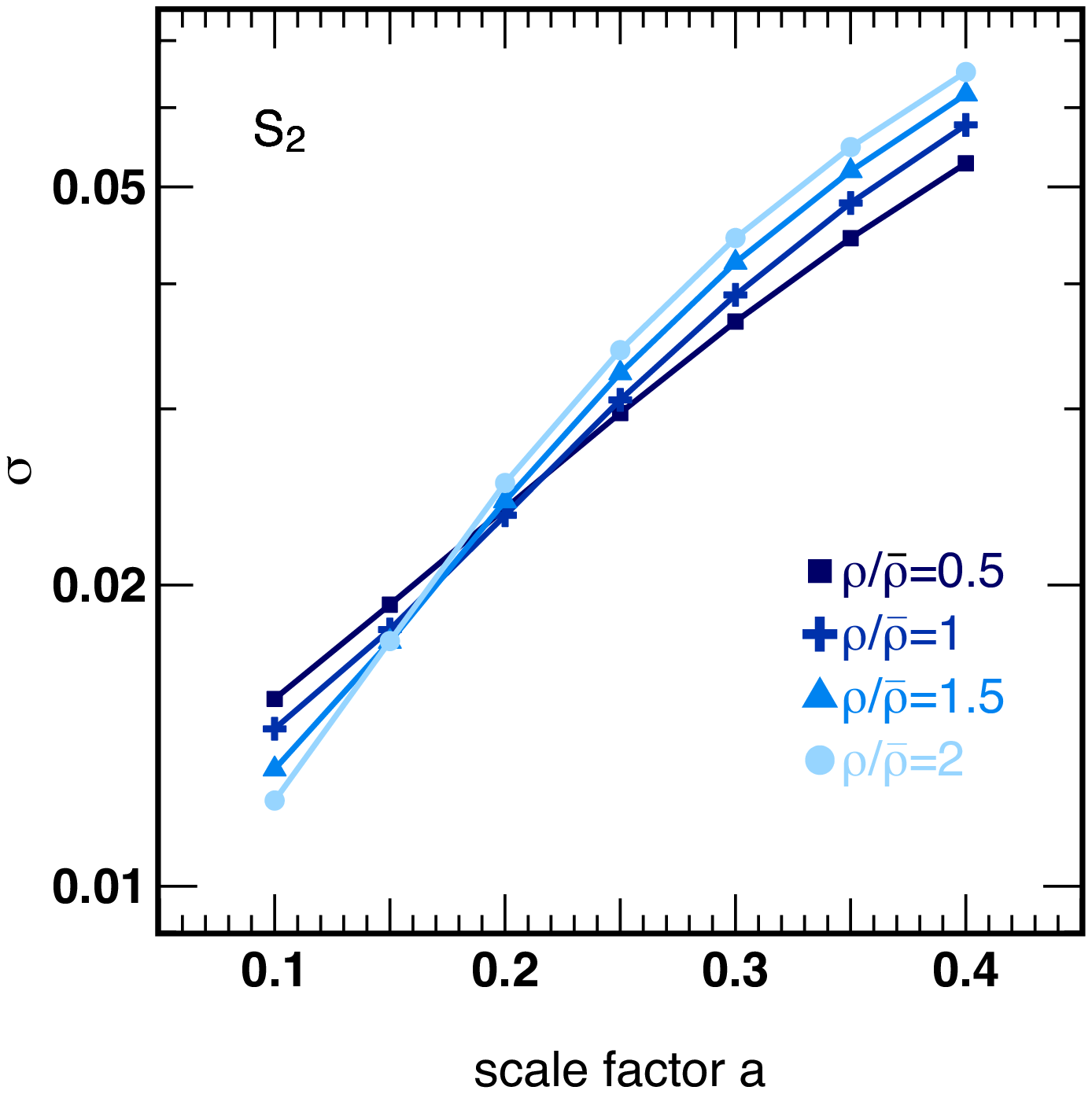}
\caption{\label{fig:density} {\it top}: Evolution of the
pixel-averaged PDF of the density values while sampling $S_{1}$ on a
$64^{3}$ grid for different values of
$a\in\left[0.1 (light),\cdots0.4(dark)\right]$ (only the pixels where
$\rho/\bar{\rho}>2$ were considered, where $\bar{\rho}$ is the cosmic
mean density). As expected the full width half
max (FWHM) of the distribution increases exponentially with the
expansion factor reflecting the chaotic behaviour of the PDF of the
density field; {\it middle}: temporal evolution of the dispersion in
the sampled density field per unit of the mean density, for sub
regions of $S_{1}$ corresponding to thresholds in overdensity
$\rho/\bar{\rho}$ of 0.5, 1, 1.5 and 2 respectively as labelled.
 The asymptotic merging
for different thresholds reflects the fact that at later times,
regions of different overdensity levels are all in the nonlinear
regime; {\it bottom}: same as middle frame but for $S_{2}$.}
\end{figure}
Specifically, Figure \ref{fig:density} presents the evolution of the
mean relative dispersion of the density, $\delta\rho/\rho$ (where
$\rho$ is the mean density of the pixel over all the realisations and
not the average density of the simulation), in
identical pixels of the different realisations of $S_{1}$ (middle
panel) and $S_{2}$ (bottom panel), considering regions where density
is greater than given thresholds.\footnote{note that the number of pixels above a given 
threshold is going to depend on redshift, but for the contrasts considered here, 
the error on the dispersion due to shot noise is always negligible, as
we have at least $8000$ particles above the highest threshold, at the
highest redshift.} 

As expected, these measurements show
that this dispersion increases with time, as can easily be seen on the
{\sl top panel } of figure \ref{fig:density}, where the PDF of
$\delta\rho/\rho$ is plotted for different values of $a$. The fact
that the growth rate of the dispersion increases when considering
regions of higher densities may be explained by the higher level of
nonlinearity of the evolution of matter distribution in these
regions. In fact, in denser regions, the evolution becomes non-linear
earlier, which favors the development of chaos.{ But at later times,
non linearities have had time to develop at all considered overdensity
levels, which explains the asymptotic merging of the curves. } The
exponential growth of the dispersions demonstrates that the evolution
is chaotic as defined in the introduction and allows for the
computation of Lyapunov exponents, $\lambda_{P}$, as the rate of
change of the logarithm of the average relative density fluctuation as
a function of the scale factor.

\begin{figure}
\includegraphics[width=0.9\columnwidth]{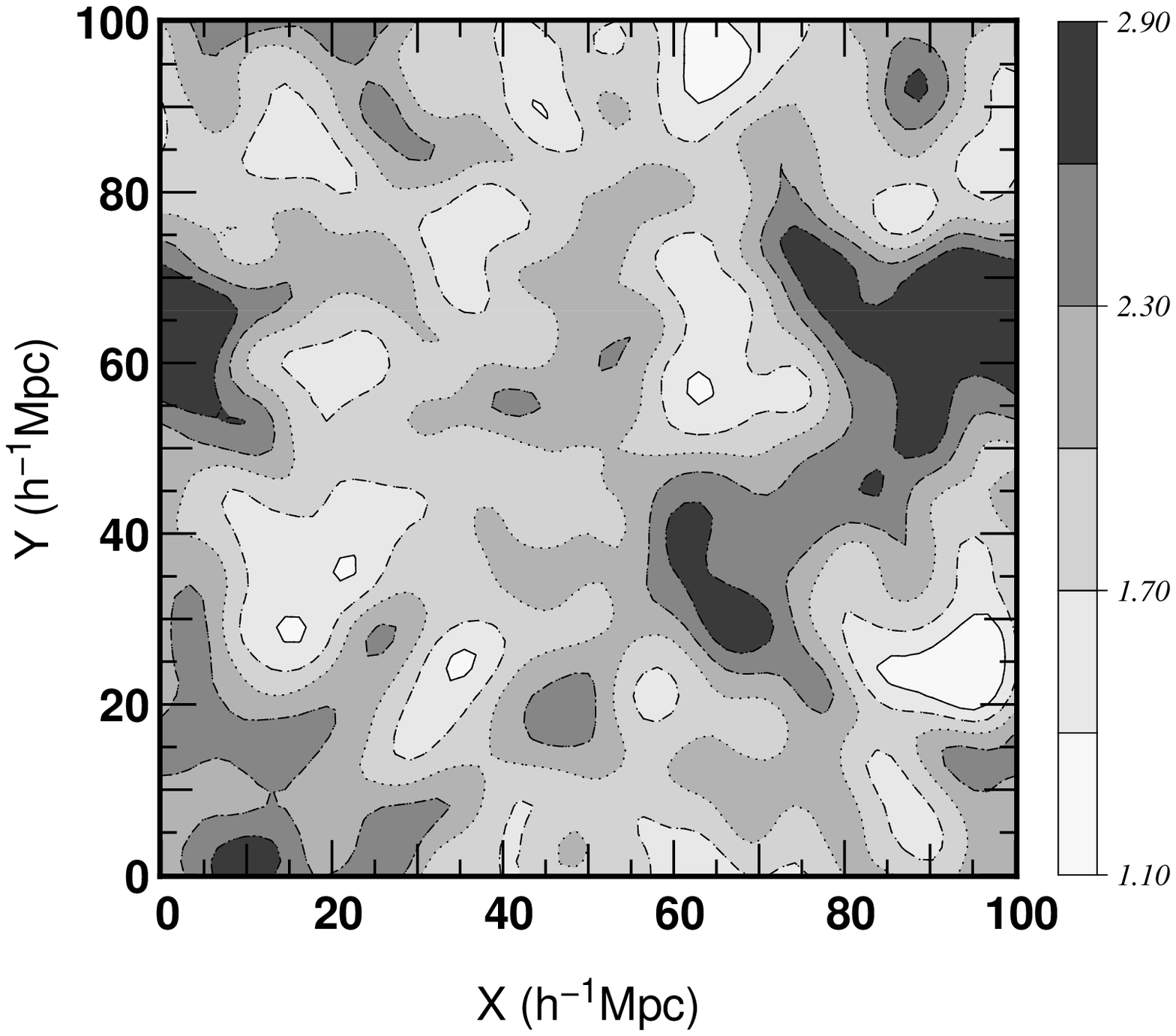}
\includegraphics[width=0.9\columnwidth]{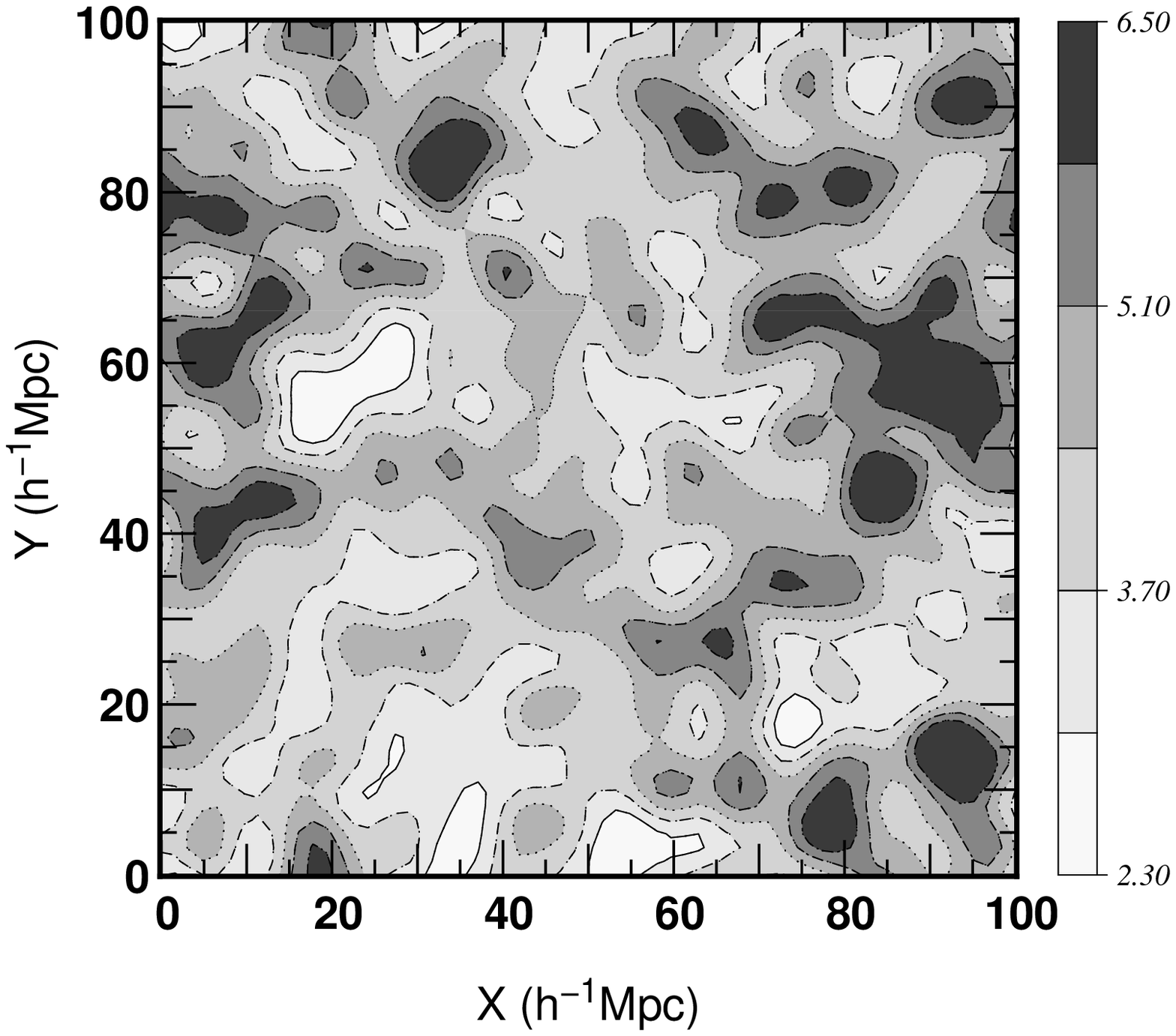}
\caption{\label{fig:pixel}Logarithm of the projected density of the
 pixels (\textsl{top} frame) and their associated Lyapunov exponent
 (\textsl{bottom} frame), for a projected $10h^{-1}$Mpc slice $S_{2}$
 at $a=0.35$. The comparison of the two maps emphasizes the
 correlation of the two fields: denser regions have larger Lyapunov
 exponents. On closer inspection, one may argue that larger Lyapunov
 exponents lie in the outskirts of the denser regions.  }
\end{figure}

The fact that the non-linearity in the evolution increases chaos is
illustrated by Figure \ref{fig:pixel}, where maps of the average
density ({\sl top panel}) and the corresponding Lyapunov exponent
$\lambda_P$ (bottom panel) are plotted.
Each map represents the projection of a $10h^{-1}$Mpc slice from a
sample of $S_{2}$ at $a=0.35$. The correlation between the two maps
confirms the dependence of chaos on over density (see also the
projections of different realisations of the same halos on Figure
\ref{fig:halo1}, where substructures are clearly different even though
the shape of main halos remains mostly the same). These results must
nonetheless be interpreted with care as the use of a finite sampling
grid may bias the measurements. Indeed, considering higher density
regions amounts to considering smaller scale regions, of order the
size of the grid pixels ($\approx 1.5h^{-1}$ Mpc$^3$), which may
affect the measured value of $\lambda_{P}$.
%

\subsection{Chaos transition scale}\label{sub:trans}

\begin{figure}
\includegraphics[width=0.9\columnwidth]{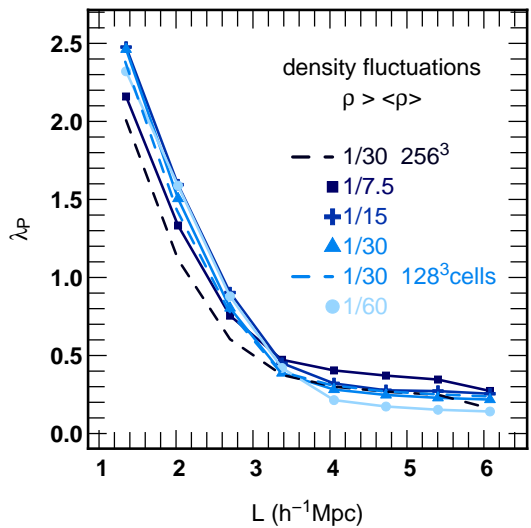}
\caption{Evolution of the average Lyapunov exponent of the pixel
 density fluctuations, $\lambda_P$, as a function of the smoothing
 length $L$ for regions where $\rho/\bar{\rho}>1$ and for different
 amplitudes of the initial perturbations, $A$ (expressed as a fraction
 of the initial dispersion amplitude) measured in the set $S_{1}$.
 The top dark dashed line  corresponds to the set $S_{2}$ for $A=1/30$.  The
 sharp transition near $L_{\rm smooth}\approx 3.5 {\rm Mpc}/h$ is exhibited in
 both resolutions.  The perturbation amplitude does not affect the
 result significantly.  {The difference in time of measurement for the
 $S_{2}$ curve (slightly earlier than the freeze-out time around
 $z\sim1$) may explain small the difference in the corresponding non
 linear scale.  The  bottom light dashed line
  corresponds  also to  $A=1/30$ in $S_{1}$ but 
 measured  on a $128^{3}$  grid; it shows that the exponent is not 
 sensitive to the sampling resolution.
  } %
\label{fig:lambdaVSsmooth}}
\end{figure}

Transition to chaotic behaviour of the density field that started with
linear evolution is fundamentally linked to the development of the
nonlinearity.  Since different scales enter nonlinear regime at
different epochs, one expects that at a given time there exist a
transition scale, $L_c$, below which variation of the density in
pixels of the sampled field is clearly chaotic. Figure
\ref{fig:lambdaVSsmooth} presents the behaviour of the average value
of $\lambda_P$ for different perturbation amplitudes $A$, as a
function of the scale $L$. These measurements are derived by computing
the average Lyapunov exponents in pixels on the sampled maps shown in
figure \ref{fig:pixel}, smoothed using a Gaussian kernel of FWHM $L$,
and considering only the overdense regions ($\rho/\bar\rho>1$).
Density is computed by making a histogram of particles in the grid
using the NGP method, and by smoothing it with a Gaussian kernel afterwards.
 The measurements are performed at the present time, $a=1$ in $S_1$
simulation and at $a=0.4$ for $S_2$ set.

The plot demonstrates a rather sharp transition to chaotic behaviour at
scales below the critical smoothing length $L_c \simeq 3.5 h^{-1}$ Mpc
with Lyapunov exponent increasing for ever smaller scales, whereas on
larger scales the Lyapunov exponent is small and constant. This
behaviour is indicative of the $\Lambda$CDM background cosmology of
the standard model. Indeed, in the pure CDM cosmology with the
critical density of the matter, the gravitational clustering would
have continued to escalate to present time and one expect to see
Lyapunov exponent falling smoothly to $L \simeq 8h^{-1}$Mpc, the
present-day nonlinear scale
\footnote{The nonlinear scale is usually defined with top-hat
smoothing as $\sigma^2(R_\mathrm{TH})=1$. The FWHM of the Gaussian
smoothing filter $L$ that we use gives similar variance to the top-hat
filter at $R_\mathrm{TH} \approx 0.9 L$. Our simulations are
normalized to $\sigma(8 h^{-1} {\rm Mpc})=0.92$ which at $a=1$
corresponds to nonlinear scale $R_\mathrm{TH} \approx 7.2 h^{-1} $Mpc,
i.e $L \approx 8 h^{-1} $Mpc.}.
  In contrast, in $\Lambda$CDM
cosmology the hierarchical clustering saturates when the dark energy
begins to accelerate the expansion of the Universe.  Numerical
simulations show that in the standard $\Lambda$CDM model the
clustering largely ceases by $z \sim 1$ (\citep{Hatton}). The non-linear scale at this
redshift is $L=3.7 h^{-1} $Mpc, which corresponds to the mass scale $M
\approx 2 \times 10^{13} M_\odot$. The halos of smaller mass collapse
en masse at earlier times passing by $z=1$ through a period of
hierarchical mergers with similar-mass halos as well as accretion that
contributes to the formation of the chaotic features. Whereas the
larger overdense patches, even the rare ones that turned around by $z
\sim 1$ and will collapse by the present time, evolve in a quiescent
environment of frozen hierarchy \citep{VdB, aubert2}. This argues for $L \approx 3.7 h^{-1}
$Mpc providing the fixed critical length between chaotic and regular
regimes for all $z < 1$, which is in general agreement with our
measurements.

\section{Lagrangian exponents}\label{sec:lagr}

In the previous section, we studied the development of chaos in the
density field of cosmological simulations. We measured the evolution
of the variance of this density field on a grid (i.e. at peculiar {\em
Eulerian} locations) and showed that chaos tends to be more pronounced
in higher density regions as well as on smaller scales. Let us now
focus instead on {\em Lagrangian} properties of peculiar objects with
a physical significance such as dark matter halos or filaments.

\subsection{Inter skeleton distance}\label{sec:skel}
Filaments correspond to a central feature of the large scale
 distribution of matter: large void regions are surrounded by a
 filamentary web linking haloes together. Studying the properties of
 the filaments isn't an easy thing and one first needs to find a way
 of extracting their location from a simulation. The skeleton gives a
 mathematical definition of the filaments as the locus where, starting
 from the filament type saddle points (i.e.  those where only one
 eigenvalue of the Hessian is positive), one reaches a local maximum
 of the field by following the gradient. This is equivalent to solving
 the equation:
\begin{equation}
\frac{d \mathbf{x}}{dt}\equiv \mathbf{ v}=\nabla\rho\,,\label{eq:skel}
\end{equation}
for $\mathbf x$, where $\rho(\mathbf{x})$ is the density field,
$\nabla\rho$ its gradient, and $\mathbf x$ the position. Although
apparently simple, solving this equation is quite difficult which is
why a local approximation was introduced in (\cite{skel1,skel2}): the
\emph{local} skeleton. One can show that, up to a second order
approximation, solving Equation (\ref{eq:skel}) is equivalent to
finding the points in the field where the gradient is an eigenvector
of the Hessian matrix together with a constraint on the sign of its
eigenvalues. This approach leads to a system of two differential
equations, solved by finding the intersection of two isodensity
surfaces of some function of the density field and its first and
second derivatives. This procedure is very robust and allows for a
fair detection of the dark matter filaments.
\begin{figure}
\includegraphics[scale=0.5]{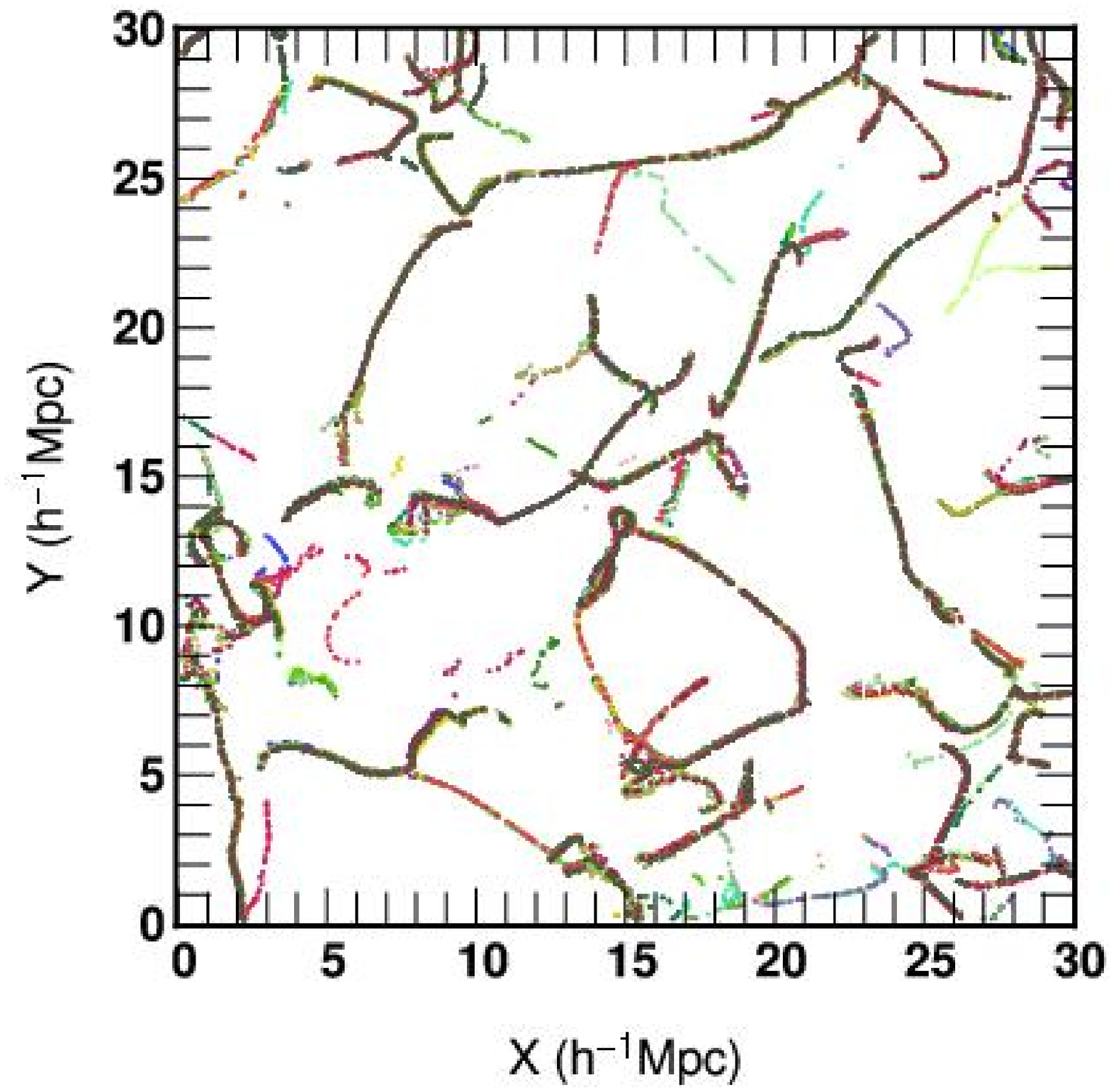} 
\includegraphics[scale=0.5]{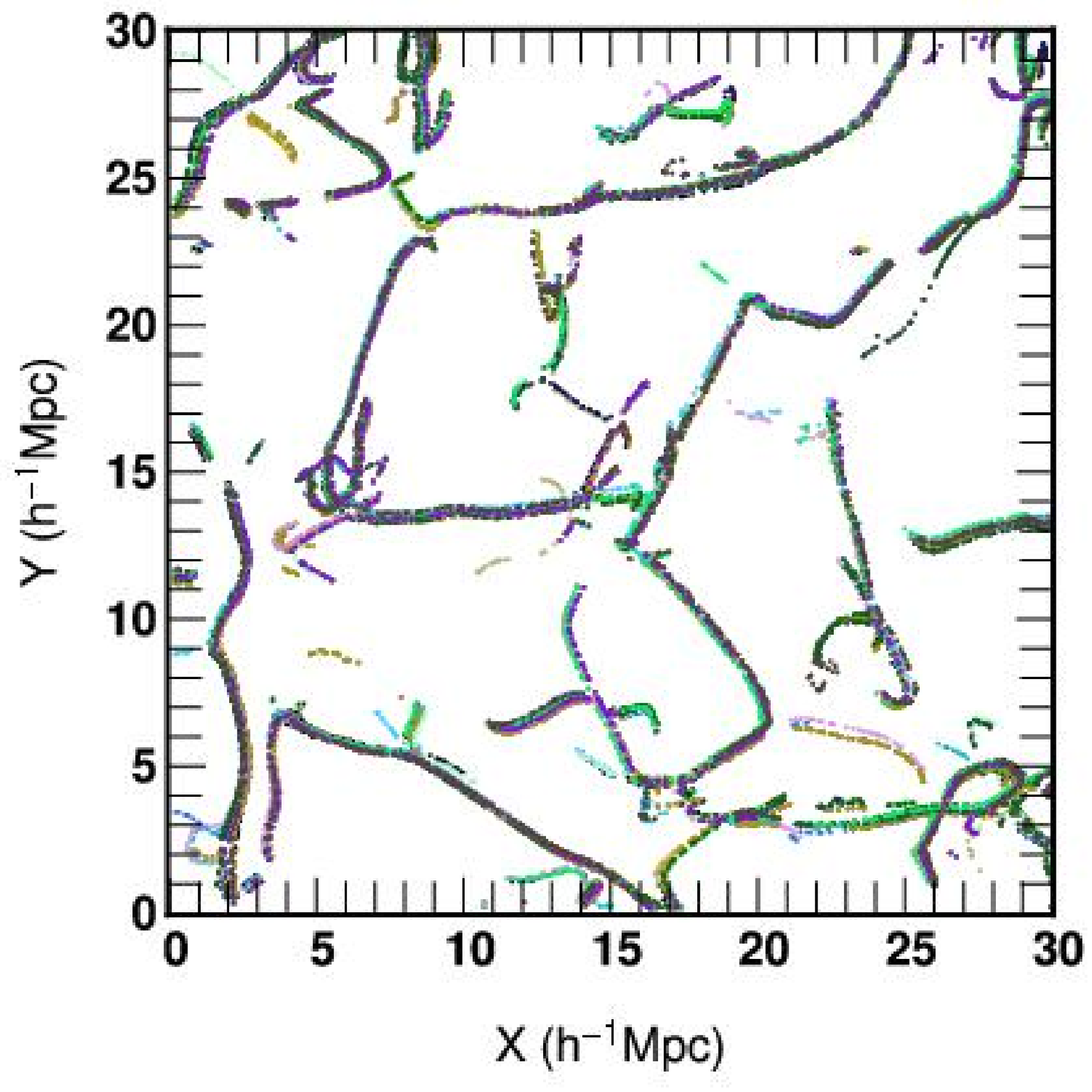}
\caption{\label{fig:skels} The \emph{local} skeletons of the different
realisations of $S_{2}$, computed at $a(t)=0.1$ (\textsl{top}) and
$a(t)=0.4$ (\textsl{bottom}) and for a smoothing length
$L_{0}=1.2h^{-1}$Mpc. Each figure corresponds to the projection of a
$10h^{-1}$ Mpc thick slice. Each color represents a different
realisation of the simulation, the color coding is not consistant
between the top and the bottom panels.
The dispersion in the position of the
skeletons appears to have grown from $a(t)=0.1$ to $a(t)=0.4$.}
\end{figure}
Figure \ref{fig:skels}
displays the skeleton of different realisations of $S_{2}$ at
$a(t)=0.1$ (\emph{top}) and $a(t)=0.4$ (\emph{bottom}).
Note that the dispersion of the skeleton location has
increased with the scale factor.
Using the method described in (\cite{skel1,skel2}), the \emph{local}
skeleton provides a list of small segments. In order to measure the
distance between two skeletons, for each segment, the distance to the
closest segment in the other skeleton is computed leading to the PDF
of this distribution. The mean distance between the two skeletons is
set to the position of the first mode of their inter-distance PDF (See
also \cite{caucci}).  We then define a mean inter-skeleton distance
among all the realisations within a set as the arithmetic average of
their pairwise distances.  This means that the normalized
inter-skeleton distance, $\left\langle D\right\rangle/L_{0}$, is our
measure of the dispersion in the skeleton location. It is a Lagrangian
property since it follows the flow.  Its evolution as a function of
the scale factor is plotted on Figure \ref{fig:regPDF} for different
smoothing lengths $L_{0}$, for $S_2$ (\emph{top}) and $S_1$
(\emph{bottom}).  The smoothing operation is achieved, as previously,
by convolving the density field with a Gaussian kernel of FWHM
$L_{0}$, ranging from $L_{0}=1.2\,h^{-1}$Mpc (3 pixels) up to
$L_{0}=3.5 \,h^{-1}$Mpc (9 pixels).
\begin{figure}
\includegraphics[width=0.9\columnwidth]{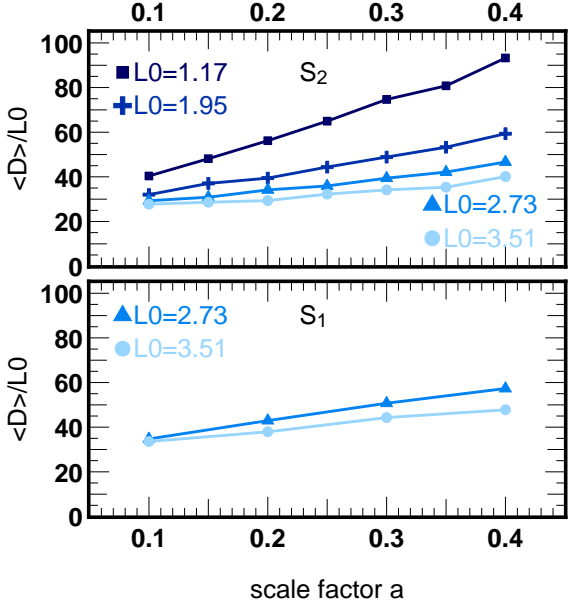}
\caption{\label{fig:regPDF} The normalized mean distance,
$\left\langle D\right\rangle/L_{0}$, between the skeletons of $S_{2}$
(\emph{top}) and $S_{1}$ (\emph{bottom}) as a function of the scale
factor and for different values of the smoothing length, $L_{0}$ in $h^{-1}$ Mpc.
Because of the lack of accuracy at smaller scales, only the larger
smoothing lengths are represented for $S_{1}$.   At these resolutions, the two sets agree.
The cosmic web
dispersion clearly evolves linearly with time, confirming that chaos
is linked to non-linearities.
}
\end{figure}
It is clear that whatever the smoothing scale or the resolution used,
the evolution of the dispersion is linear with the scale factor.  A
shift in the skeleton of the initial conditions will evolve linearly
with time and not exponentially: the skeleton at present time won't be
affected very much. There is no chaotic drift of the position of the
skeleton and thus no chaos in the evolution of the cosmic web.  Note
nonetheless that the smaller the smoothing length, the stronger the
increase of $\left\langle D\right\rangle/L_{0}$. This implies that
smaller scales are more sensitive to initials conditions, which is
confirmed by the fact that $\left\langle D\right\rangle/L_{0}$ is
larger for lower values of $L_{0}$, whatever the value of $a$.

\subsection{Positions of halos}\label{sub:pos}
\begin{figure}
\includegraphics[width=0.95\columnwidth]{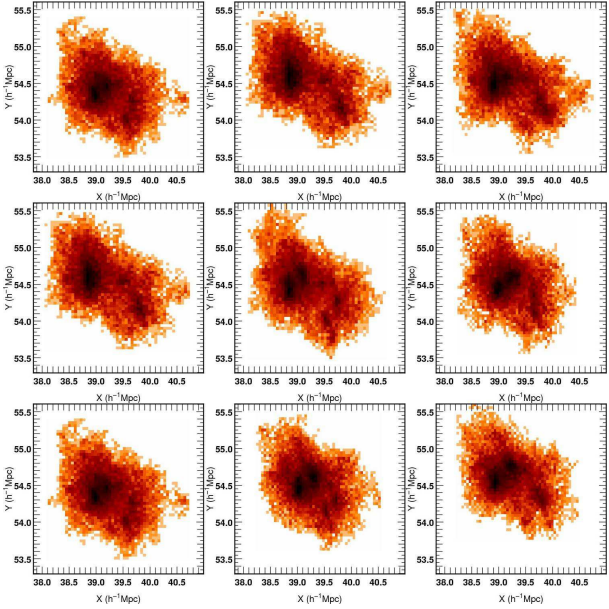}
\caption{\label{fig:halo1} Heaviest cluster (in the X-Y plane in
Mpc/h) of 9 realisations of $S_{2}$ at $z=1.5$. The position and the
global shape of the halo does not change from one simulation to
another, but the substructures are quite different; this is confirmed 
via  automated substructure identification using \emph{ADAPTAHOP}.}
\end{figure}
Turning to stochasticity on smaller scales in a Lagrangian framework
(i.e. ignoring the absolute shift in position relative to a fixed
frame), let us define a matching procedure to identify structures in
different runs. haloes are first identified using the FOF algorithm
(\cite{fof2,fof1}) with a percolation length of $0.25h^{-1}$Mpc for
$S_{1}$ and $0.5h^{-1}$Mpc for $S_{2}$ corresponding to
$0.2\times$ mean interparticular distance. 
In order to tag different FOF haloes in different
realisations as counterparts, all particles of a given halo are
matched in another realisation using their initial index
(Figure~\ref{fig:halo1}). The halo of the other simulation containing
most of these particles is tagged as its counterpart. The procedure is
carried over all pairs of simulations, allowing the measurement of the
variation in the halo properties like their spins, their positions,
their velocity dispersion tensors or their masses.

As shown on Figure \ref{fig:halo1}, the haloes locations seem
relatively insensitive to small changes in the initial conditions. The
evolution of the mean distance between a halo in a given simulation
and the same halo in another realisation is linear, as for the
skeleton, which confirms the first impressions: no chaos is observed
at linear scales and so $\lambda_Q$, the Lyapunov exponent of the
inter halo distance, is null. But the most interesting results
involve the substructures. The halo pictured in Figure
\ref{fig:halo1} is a good example of the generic behaviour. The number
of substructures changes from one realisation to another (here, 1 or 2
substructure(s)) and their positions also differ. These
results are confirmed by an automated detection of the substructures using
\emph{ADAPTAHOP} (\cite{aubert}). Both the locations
and the number of substructures are possibly subject to chaos, but the
lack of a cross identification procedure makes it difficult to
quantify it and is somewhat beyond the resolution of these sets of
simulations (Section~\ref{sub:connexity} addresses this problem for
the FOF halos). This trend confirms quantitatively the findings of
section \ref{sub:density} from the point of view of Eulerian
estimators which are sensitive to the detailed extension of the
distribution of matter within halos: denser regions were found to be
chaotic, and will be addressed in more details in Section~
\ref{sub:Spin-Orientation} in terms of halo density and velocity
moments.

\subsection{Connexity and mass of clusters}\label{sub:connexity}
The connexity of haloes can be defined as follows: considering the
$p^{th}$halo, $H_{p}^{r}$, in the $r^{th}$ realisation, its particles
are spanned amongst $n$ haloes in the $r'^{th}$ realisation, and a
fraction $f_{pk}^{rr^{\prime}}$ ($k\in{1,..,n}$) of them belong to a
halo $k$ among $n$ in the $r^{\prime th}$ realisation. Hence, its
relative connexity $C_{p}^{rr^{\prime}}$ can be defined as:
\begin{equation}
C_{p}^{rr^{\prime}}=\sum_{i=1}^{n}i\,\left(\prod_{j=1}^{i}\sum_{k=j}^{n}f_{pk}^{rr^{\prime}}\right)\,,
\label{eq:defC}
\end{equation}
where by construction $C_{p}^{rr^{\prime}}$is equal to one if both
haloes are identical in realisations $r$ and $r'$;
$C_{p}^{rr^{\prime}}$ is equal to $n$ if the halo $p$ splits into $n$
haloes with equal fractions $f_{pk}^{rr^{\prime}}=1/n$ in realisation
$r^{\prime}$, while preserving continuity when the values of
$f_{pk}^{rr^{\prime}}$differ, see Appendix~A.  The mean connexity,
$C,$ is obtained by averaging over all haloes containing more than
$100$ particles in every possible combinations of realisations and is
a measure of the dispersion of the particles.

As shown on Figure \ref{fig:connexity}, $C$ increases with the scale
factor, ranging from $1.17$ (i.e., statistically, $~90$\% of the
particles belong to a unique halo in other realisations) to $1.37$
($~85$\%) from $a=0.2$ up to $a=1.0$. The connexity clearly does not
vary exponentially with the scale factor : there are statistically no
halo fission during evolution (thanks to the efficiency of dynamical
friction). Moreover, two haloes marginally linked by FOF would almost
always end up merging sooner or later.  More and more haloes merge at
different times in different realisations which is in part due to the
fact that some threshold is involved in the FOF algorithm: a precise
linking length has to be chosen, inducing the possibility that small
changes in particles position can induce significant changes in halo
merging time (according to the FOF definition of a halo). At later
times ($a>0.7$), the connexity reaches a plateau, which suggests that
when haloes are massive enough ($M\geq M_{\rm c}$, see Section~\ref{sub:Spin-Orientation} below), they become insensitive to the merging
of lighter ones, since equal mass merging rarely occur below $z=1$.
The analysis of the masses of the haloes shows that there is no
sweeping change and so, no obvious evolution of the haloes mass
distribution: the associated Lyapunov exponent, $\lambda_M$, is
null. The number of different particles increases with time but the
missing particles are replaced by new particles. Thus, the mass stays
quite constant even as the connexity increases.  It follows that the
mass function extracted from the N-body simulations are found to be
quite robust with respect to changes in the initial conditions.
Although the masses of the haloes are similar in different
realisations, some of the particles which compose them may be
different, which may generate differences in the physical properties
of the haloes. The substructures are different (Fig. \ref{fig:halo1})
in their numbers and positions, which is responsible for the Eulerian
chaos found in Section~\ref{sub:trans}.  Let us now re-explore this in
a Lagrangian framework.
\begin{figure}
\includegraphics[width=0.9\columnwidth]{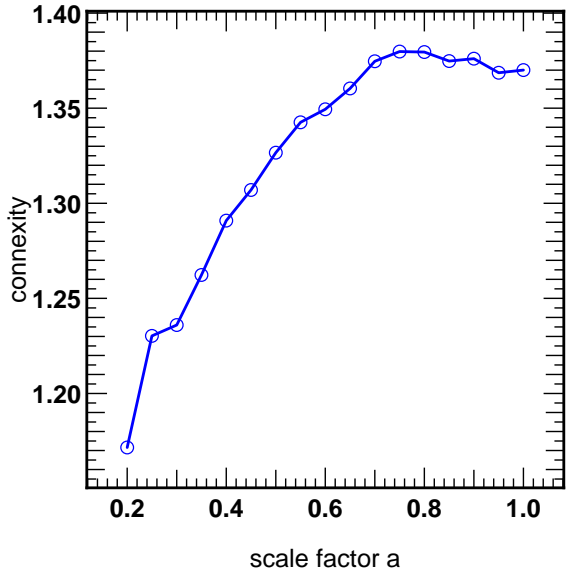}
\caption{\label{fig:connexity} The haloes average connexity computed
over all the realisations of $S_{1}$ as a function of the scale
factor. While the connexity is not subject to chaos, its value
increases with time. This result can be understood through the
difference in merging time of the haloes.}
\end{figure}
\subsection{Spin Orientation of  clusters}\label{sub:Spin-Orientation}
The influence of chaos on the spin of haloes is estimated by computing
the cosine of the angle, $\theta_{pq}$, between the spins
$\boldsymbol{J}_{p}$ and $\boldsymbol{J}_{q}$ of corresponding haloes
in two different realisations p and q:
\begin{equation}
\cos(\theta_{pq})=\frac{\boldsymbol{J}_{p}\cdot\boldsymbol{J}_{q}}{\left\Vert
\boldsymbol{J}_{p}\right\Vert \left\Vert
\boldsymbol{J}_{q}\right\Vert }.
\end{equation}
For every bin of mass, a measure of the dispersion, $\sigma$, of the
orientation is given by the average angle:\footnote{the estimator of
the dispersion, Eq.~(\ref{eq:sigangle}) is robust since weighting the
sum by the spin parameter yields the same results.}
\begin{equation}
\sigma=\arccos\left(\frac{1}{N_{c}}\sum_{i=1}^{N_{c}}\cos\theta_{pq}\right),\label{eq:sigangle}
\end{equation}
where the sum is over all the $N_{c}$ possible pairwise combinations
of realisations. Note that only bins of masses containing more than 30
haloes have been retained.

Figure \ref{fig:spin} displays the exponential growth of this
dispersion with time, and shows that the precise value of its
associated Lyapunov exponent $\lambda_{\sigma}$ depends on the
selected bin of mass. It also shows that the exponent does not seem to be 
sensitive to shot noise, as its value is left unchanged when resolution is 
increased between $S_{1}$ and $S_{2}$.
 
Also, as it is seen in Fig.\ref{fig:halo1}, the detailed distribution
of satellites within a given cluster varies from one realisation to
another; the angular momentum orientation (in contrast to say, its
modulus or the halo mass) is quite sensitive to the outer region of
the distribution.  Recall that the spin parameter (i.e. the
$\Lambda=J/(\sqrt{2} M V_{200} R_{200})$ (\cite{spin}, \cite{aubert})
of a halo displays no chaotic behaviour. It stays quite constant from
a simulation to another and the evolution of its dispersion is not
exponential.

 The measured Lyapunov exponent ranges from 0 to 0.3. The value of the
mean dispersion of the orientation of the spin for heavier haloes is
about 35 degrees ($=\exp(3.55)$).
Globally this suggests that the orientation of the spin varies with
  the tidal field, which in turn depends on the relative position of
  structures within the environment of the halo.

 For lighter haloes, the measured value of $\lambda_{\sigma}$ is
higher than for heavier ones (Figure \ref{fig:spin}) which may be
partly explained by the fact that a slight change in a few clumps
within the haloes has a larger influence on its spin when they
represent a significant fraction of it.  Faltenbacher \&
al. (\cite{spinmerging}) showed that if the lightest halo has a mass
less than $10\%$ of the mass of the larger halo, the orientation of
the resulting post merging halo will remain statistically the same. In
contrast, if its mass is greater than $20\%$ of the more massive halo,
the final orientation of the merged halo depends on the speed vector
of the two progenitors. Our results corroborate well their finding
since the lightest haloes that are formed by merging of two
substructures of comparable masses have chaotic spins (substructures
being chaotic, see Section~\ref{sec:euler} and
Figure~\ref{fig:halo1}), while heavier ones have spin that are relatively
stable with time (they only merge with much smaller haloes).  
\begin{figure}
\includegraphics[width=1.25\columnwidth]{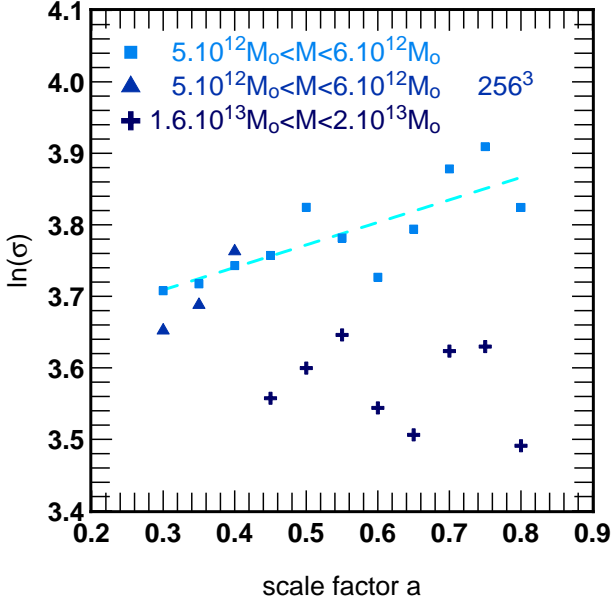}
\caption{\label{fig:spin}Logarithm of the dispersion of the angle
between the spin of one halo of $S_{1}$ as a function of the scale
factor. Results are computed for different ranges of masses, $5\,
10^{12}$M$_{\odot}$<M<$6 \, 10^{12}$M$_{\odot}$ (\textsl{top}) and
$1.6\, 10^{13}$M$_{\odot}$<M<$2\, 10^{13}$M$_{\odot}$
(\textsl{bottom}).  
For heavier halos the Lyapunov exponent vanishes. The triangles correspond to the first class of lighter mass, but measured in $S_{2}$; the exponent remains unchanged which suggests that particle
shot noise is not an issue.
 }
\end{figure}
It emerges from these measurements that there is a critical mass,
$M_{c}$, above which chaotic behaviour disappears.  Haloes heavier
than this mass are too heavy to feel the influence of incoming clumps
and their spins are clearly defined. They are therefore not subject to
chaos and their Lyapunov exponents are null at the one sigma level,
 in contrast to lighter 
ones whose spin are sensitive to the initial conditions and whose
Lyapunov exponents are positive.

As for the critical smoothing length (see Section \ref{sec:euler}), we
can study the evolution of this critical mass as a function of the
amplitude, $A$, of the perturbations. Figure \ref{fig:massecrit} shows
this evolution.  A good fit of this transition mass is given by $M_{c}
= 2\, 10^{13}\, M_{\odot} A^{0.15}$. The higher the amplitude of the
perturbations, the higher the required time for haloes to have a spin
clearly defined. Consequently, the critical mass increases with the
perturbation amplitude, and haloes that can be considered stable are
heavier. Note that it means that the spin is constant with time but not
very reliable since its final orientation depends, in part, on the
initial conditions.
\begin{figure}
\includegraphics[width=1.1\columnwidth]{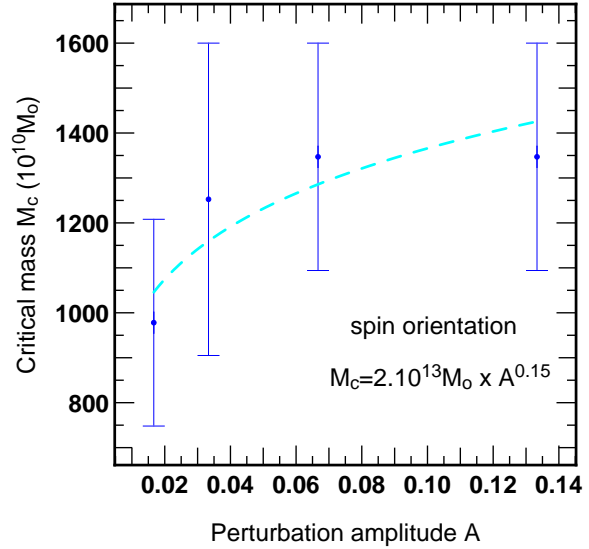}
\caption{\label{fig:massecrit} Critical mass, $M_c$, (in units of
$10^{10}$M$_{\odot}$) for the spin orientation as a function of the
amplitude of the perturbations, $A$ (in fraction of the initial
dispersion amplitude). It appears that $M_{c}= 2\,10^{13 } M_{\odot}\,
A^{0.15}$. The larger the amplitude of the perturbation, the heavier
the haloes that can be considered stable.  }
\end{figure}
\subsection{Orientation of the velocity dispersion tensor }
The orientation of the velocity dispersion tensor is also a quantity
of interest from the point of view of stochasticity since it is
related to the shape of the halo via the Virial theorem.  The
corresponding estimator involves computing the orientation of the
eigenvector $\boldsymbol{V}$ associated to the largest eigenvalue of
the velocity dispersion tensor. As for the orientation of the spin,
(Sec. \ref{sub:Spin-Orientation}) the angle $\theta_{pq}$ between the
eigenvector $\boldsymbol{V}_{p}$ of the halo in simulation $p$ and its
corresponding eigenvector $\boldsymbol{V}_{q}$ in a simulation $q$ is
computed as:
\begin{equation}
\cos(\theta_{pq})=\frac{\boldsymbol{V}_{p}.\boldsymbol{V}_{q}}{\left\Vert
\boldsymbol{V}_{p}\right\Vert \left\Vert
\boldsymbol{V}_{q}\right\Vert }.
\end{equation}
For every bin of mass, a measure of the dispersion, $\sigma$, is also
given by  Equation (\ref{eq:sigangle}).
%
 Once again, only bins of masses containing more than 30
haloes were considered.
 As shown in Figure \ref{fig:vecpropre} this estimate is consistent
with the exponents of the orientation of the spin: only the lightest
masses are sensitive to the initial conditions, while the dispersion
of the orientation for the heavier masses is constant (about 40
degrees). The measured Lyapunov exponent, $\lambda_T$, ranges from 0
up to 0.65 . These results corroborate well those for the spin axis,
given that its orientation follows the third eigenvector of the
dispersion matrix (i.e.  the axis along which dispersion is the
smallest).
\cite{spinmerging} showed that the orientation of the principal axis
of the halo is correlated with the vector linking the two mergers
(i.e. their relative positions) particularly in the case where one of the
mergers has a mass smaller than $10\%$ of the second one.  The chaos
found at small scales (substructures scale) is once again responsible
for the chaos in these geometrical properties of haloes since changing
the initial conditions amounts to changing the relative positions of
the substructures (see sec.\ref{sub:density}) and thus to changing the
orientation of the resulting halo's dispersion tensor.  As for the
spin, there is evidence of a critical mass above which chaotic
evolution disappears: more massive haloes only merge with lighter ones
that do not affect their global properties.

\begin{figure}
\includegraphics[width=1.3\columnwidth]{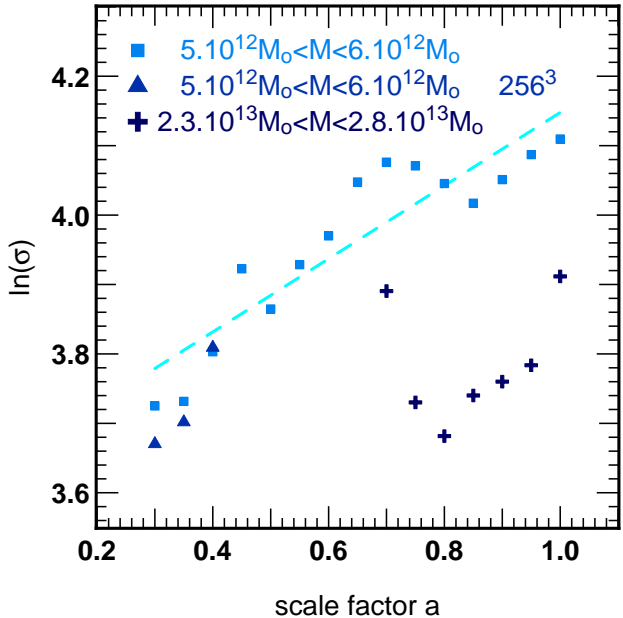}
\caption{\label{fig:vecpropre}Logarithm of the dispersion of the angle
between the first eigenvector of the velocity dispersion tensor of the
haloes of $S_1$, as a function of the scale factor. Results are
computed for different ranges of masses:
$5\,10^{12}$M$_{\odot}$<M<$6\,10^{12}$M$_{\odot}$ ({\it top}) and $2.3
\,10^{13}$M$_{\odot}$<M<$2.8\,10^{13}$M$_{\odot}$ ({\it bottom}).  The
less massive haloes are more sensitive to the initial conditions, the
average angle being constant for the heavier ones, about a value of $40$
degrees.  The triangles correspond again to the $S_{2}$ set and shows not difference.
The Lyapunov exponent, $\lambda_T$ ranges from $0$ to $0.65$.  }
\end{figure}
\section{Conclusion  and discussion}\label{sec:Conclusion}


Let us first emphasize here again that the term chaos is used in this paper in
the loose sense, as the age of the universe does not allow for many
e-foldings on larger scales.  Table~\ref{tab:exposants} summarizes the
different Lyapunov exponents computed in this paper.  As shown in
section \ref{sub:density} (Figure~\ref{fig:lambdaVSsmooth}), chaos
appears below a critical scale which corresponds roughly to cluster
scales.
The higher the density, the more chaotic is the corresponding region.  We
also found that both Lagrangian and Eulerian measurements are
consistent: super clusters and filaments, whose dynamics is
globally linear (large scale structures), are not stochastic: a shift
in the initial conditions will increase linearly with the scale
factor. By contrast, the distributions of substructures within
clusters, whose characteristic size is smaller than $\sim 3.5 h^{-1}$Mpc,
are governed by non-linear dynamics and may undergo a stochastic
evolution for some observables.

Nevertheless, this chaos at substructures scale does not occur
for all physical characteristics of the cluster's halo. A main
fraction of particles remains in the same halo from one realisation to
another, while a difference arises (in part) from the delay in merging
times of the substructures. These timing effects are however averaged
out yielding, to first order, a constant halo mass. It follows
that the mass function derived from a simulation is quite consistent
from one realisation to another. Similarly, the dispersion of the
amplitude of the total spin of haloes does not increase exponentially
with time.

The mass of a given halo is an integrated quantity which does not
trace which specific particle entered the FOF halo; similarly, the
spin parameter is also an adiabatic invariant, and the trace of the
dispersion tensor will relax rapidly to its virial expectation (which
is mass dependent) in a few short dynamical times; in contrast, the
spin orientation or the orientation of the dispersion tensor will
depend precisely on the orientation of velocities of the entering
particles and has no direct relation to the mass of the halo; it also
reflects the initial environment of the proto halo. For instance it
has been shown in \cite{skel2}, \cite{aubert} that the halos
preferentially anti-align their spin with the axis of the filament in
which they are embedded, while we have shown in Section~\ref{sec:skel}
that the filament's locus was not stochastic.

It is possible to recast these interpretations in the context of the
peak-patch (\cite{peakpatch}) description of haloes. In this
framework, massive haloes correspond to large quasi spherical patches
around density peaks, which non-linear evolution will decouple from
their neighbouring large patches thanks to the cosmic acceleration
below $z \sim 1$.  Conversely, small haloes correspond to small
typically aspherical peak-patches, and will acquire tidal torques
early on which depend specifically on the detailed white noise
realisation (which fixes the shape of the peak-patches).  In the tidal
torque theory, the mass and the spin parameter are essentially
integral functions over the volume of these patches, hence will not
depend on the initial perturbations, whereas the spin orientation
itself is sensitive to these perturbations, at least at the lower end
of the mass spectrum.  This is consistent with the low scatter
relationship between the spin parameter and the mass (\cite{aubert}).

Thanks to angular momentum leverage, the orientation of haloes is
itself affected by stochasticity mostly at small scales, a result which seems 
insensitive to shot noise as the lyuponov exponents are consistent between sets $S_{1}$ and $S_{2}$. 
 In fact, as
long as the haloes merging together have similar sizes (masses), the
orientation of both spin and velocity dispersion tensors is determined
by the relative positions and velocities of the two mergers, whose
dynamics is non-linear and whose characteristic size is below the
critical scale (Figure~\ref{fig:lambdaVSsmooth}). These results seem
robust with respect to resolution.

When the halo is formed and well-isolated by cosmic acceleration, it
merges only with satellites/substructures whose masses represent a
small fraction of the host's mass. Consequently, their orientations
are globally preserved after merging, and thus, the chaotic behaviour
stops and the dispersion in the orientation remains at the same level
(i.e. the resulting average angle is unchanged).  Hence, a critical
mass can be defined as the mass above which this chaotic behaviour of
the orientation stops.  The measured value of this critical mass,
$M_{c} = 2 \, 10^{13} M_{\odot} A^{0.15}$, is just below the scale of
nonlinearity at $z \sim 1$ and shows weak dependence on the amplitude
of the added perturbative noise: $M_{c} = 2 \, 10^{13} M_{\odot}
A^{0.15}$. Although some slow increase of $M_{c}$ with $A$ is expected
since adding power to inhomogeneities shifts the nonlinear scale
to higher masses, the details of the dependence require further
investigation.

  This paper has concentrated on a realistic $\Lambda$CDM cosmology:
 it would also be interesting to rerun this investigation on
 scale-free power spectra to confirm that the dark energy is indeed
 responsible for the saturation of $M_{\rm c}$.  A natural extension
 of this work, clearly beyond its current scope, would also involve
 computing Lyapunov exponents for the properties of substructures
 within halos (see for instance \cite{valluri}), and parameters
 corresponding to the inner structure of halos, such as NFW
 concentration parameter, the phase space density
 $Q=\rho_{0}/\sigma^{3}$ (\cite{peirani}), the Gini index or the
 asymmetry (\cite{gini}) within the FOF.

In closing, the answer to our riddle is that chaos and non-linearities
are very strongly linked, and both occur at small scales
(substructures scales) though some non linear halo parameters (spin,
mass etc...) do not seem to be subject to chaos.  While the large scale
structures in a simulation (filaments and haloes) are quite robust
both in their locus and properties, the distribution of substructures
is more sensitive to initial conditions since their numbers and
positions vary when initial conditions vary. This in turn may prove to
be a concern when generating zoomed resimulations.

\begin{table}
  \centering
  \begin{tabular}{|c|c|c|}
    \hline
    & $\lambda$     & $\tau$(Gyear)  \\ \hline
    Pixels PDF, $\lambda_P$                                        & 0-2.5       & 3.4-$\infty$          \\ \hline		
    Velocity dispersion tensor, $\lambda_T$                        & 0-0.65     & 25-$\infty$         \\ \hline
    Spin orientation, $\lambda_{\sigma}$                           & 0.-0.31      & 75-$\infty $        \\ \hline
    Inter skeleton distance, $\lambda_S$                           & $\sim$ 0      & $\infty$       \\ \hline
    Connexity, $\lambda_C$                                         & $\sim$ 0      & $\infty$       \\ \hline
    Position of the halos and substructures, $\lambda_Q$           & $\sim$ 0      & $\infty$       \\ \hline
    Mass of the halos, $\lambda_M$                                 & $\sim$ 0      & $\infty$       \\ \hline
    Spin parameter, $\lambda_{\Lambda}$                            & $\sim$ 0      & $\infty$       \\ \hline
    Mean dispersion of velocity, $\lambda_V$                       & $\sim$ 0      & $\infty$       \\ \hline
  \end{tabular}
  \caption{Lyapunov exponents of the different observables
    studied. Interestingly, many global properties of halos do not display chaotic behaviour.
    \label{tab:exposants}}
\end{table}
%
\subsection*{Acknowledgments}
\textsl{We thank  the anonymous referee for helpful suggestions, 
D.~Aubert, S. Colombi, and R. Teyssier for fruitful
comments during the course of this work, and D.~Munro for freely
distributing his Yorick programming language and opengl interface
(available at {\em\tt http://yorick.sourceforge.net/}).  This work was
carried within the framework of the Horizon project,
\texttt{www.projet-horizon.fr}.  }

\bibliographystyle{plainnat} 
\bibliography{chaos}
\appendix\label{app:density}

\section{Connexity}\label{sec:A-connexity}

Let us consider a halo, $H$, split into $n$ parts, $P_i^n, i\le n$,
with a fraction, $f_i$, of its particles in each of them, the indices
$i$ being sorted following the decreasing values of $f_i$ ($f_i\geq
f_j$ if $i<j$). A measure $C_n$ of the connexity of $H$ should
indicate the number of clumps into which it was split, this number
does not necessarily have to be an integer, depending on the fraction of the
mass of $H$ that went into each $P_i^n$.  For instance, if $n=2$, we
want to obtain $C_2=2$ when $f_1=f_2=1/2$ and $C_2\to 1$ when $f_1\to
1$ and $f_2=1-f_1\to 0$; as the indices are sorted, $0<f_2<1/2$. So,
in this case, we could write the connexity of $H$ as: %
\begin{equation}
C_2 = 1+2f_2.
\end{equation}

Now, considering that $H$ was split into $3$ parts $P_i^3$, then
$0<(f_2+f_3)<2/3$ and $0<f_3<1/3$. So $C'_3= 2+3f_3 \to 2$ if $f_3\to
0$ and $C'_3\to 3$ when $f_3\to 1/3$. It follows that $C''_3 =
(f_2+f_3)C'_3 \to 2$ when $f_2\to 1/3$ (which implies that $f_3\to
1/3$) and that $C''_3\to0$ when $f_2\to 0$ (which implies that $f_3\to
0$ also). So
\begin{equation}
C_3=1+(f_2+f_3)(2+3f_3)
\end{equation}
has the right properties to represent the connexity of a halo split
into $3$ parts. Hence, by generalizing recursively this formula, we
obtain:
\begin{eqnarray}
C_n&=&1+(f_2+\cdots f_n)\left[2+(f_3+\cdots f_n)\left[\cdots\left[n-2+\right.\right.\right.\\&&
\left.\left.\left. (f_{n-1}+f_n)\left[(n-1)+n f_n\right]\right]\cdots\right]\right],
\end{eqnarray}
which can be developed as:
\begin{eqnarray}
C_n \!\!\!&=& \!\!\!\!1+2(f_2+\cdots f_n)+3(f_2+\cdots f_n)(f_3+\cdots f_n)+\\  &&
\cdots+n(f_2+\cdots+f_n)(f_3+\cdots+f_n)\cdots(f_n)\,,\\
&=&\sum_{i=1}^{n}i\,\left(\prod_{j=1}^{i}\sum_{k=j}^{n}f_{k}\right)\,, \label{eq:CnAp}
\end{eqnarray}
which corresponds to Eq.~(\ref{eq:defC}).  {Note that by construction,
the braket in Eq. (\ref{eq:CnAp}) is smaller than $1/i$, so that
$C_{n}$ is always smaller or equal to $n$.}

\end{document}